\documentclass[preprint,preprintnumbers,amsmath,amssymb,nofootinbib]{revtex4}
\usepackage{graphicx}
\usepackage{dcolumn}
\usepackage{bm}
\usepackage{multirow}
\usepackage[]{graphicx}
\usepackage{rotating} 
\usepackage[usenames, dvipsnames]{color}
\begin{document}
\title{\bf Low-frequency Gravitational Wave Detection \\via Double Optical Clocks in Space
}

\author{Jianfeng Su$^{1,2}$}
\thanks{These two authors contributed equally}
\email{sujf@nssc.ac.cn}

\author{Qiang Wang$^{2}$}
\thanks{These two authors contributed equally}
\email{wang@physik.uzh.ch}

\author{Qinghua Wang$^{3}$}

\author{Philippe Jetzer$^{2}$}

\address{1.National Space Science Center, Chinese Academy of Sciences, Beijing, China\\
         2.Physik Institut, University of Zurich, Winterthurerstrasse 190, 8057 Zurich, Switzerland\\
         3.Orolia Switzerland SA (Spectratime),Vauseyon 29, 2000 Neuch$\hat{a}$tel, Switzerland}

\date{\today}
\begin{abstract}

We propose a Doppler tracking system for gravitational wave detection via Double Optical Clocks in Space (DOCS). In this configuration two spacecrafts (each containing an optical clock) are launched to space for Doppler shift observations. Compared to the similar attempt of gravitational wave detection in the Cassini mission, the radio signal of DOCS that contains the relative frequency changes avoids completely noise effects due for instance to troposphere, ionosphere, ground-based antenna and transponder. Given the high stabilities of the two optical clocks (Allan deviation $\sim 4.1\times 10^{-17}$ @ 1000 s), an overall estimated sensitivity of $5 \times 10^{-19}$ could be achieved with an observation time of 2 years, and would allow to detect gravitational waves in the frequency range from $\sim 10^{-4}$ Hz to $\sim 10^{-2}$ Hz.

\end{abstract}

\maketitle

\section{Introduction}
Gravitational waves (GWs) were predicted in the theory of General Relativity (GR) more than a century ago by A. Einstein, and their basic properties can be described by solving Einstein field equations~\cite{Einstein1916,Einstein1918,Einstein1937}. A lot of efforts have been put to detect GWs experimentally in the past several decades, since the knowledge of GWs allows for a deeper understanding of the structure and evolution of the universe, as well as of the theory of GR.\\

There are mainly three types of GW signal: (1) periodic waveforms from well-modelled sources, (2) signals from stochastic backgrounds with statistical behaviour, (3) burst signals from transient sources~\cite{Moore2014}. The most interesting GW signal for this work is the first one, since it can be simplified to a sinusoid if the variance of the wave frequency of the signal over the whole duration $T$ of the observation is much smaller than a resolution bandwidth $1\slash T$~\cite{Tinto2002, Armstrong2006}.\\

Five feasible and basic methods have been adopted in order to detect GWs: (1) laser interferometer on ground~\cite{Abbott2016}, (2) Doppler tracking system~\cite{Estabrook1975}, (3) laser interferometer in space~\cite{Seoane2012,Seoane2017,Luo2016}, (4) resonant-mass gravitational waves detectors (e.g. Weber bar) ~\cite{Weber1961, Aguiar2010}, (5) pulsar timing arrays~\cite{Backer1986}. The first one aims at high-frequency GWs ($\sim 10^2$ Hz), and in 2016 the advanced Laser Interferometer Gravitational-wave Observatory (LIGO) reported the first detection of GWs due to merging of two black holes~\cite{Abbott2016}. The second and third methods are expected to be sensitive at low-frequency signals ($10^{-4}$ Hz $\sim10^{-2}$ Hz). Among the projects of the low-frequency GW detection, the Cassini GW experiment (Doppler tracking system) by NASA finished its mission in September of 2017, but no detection evidence has been reported~\cite{Cassini}. The space-based mission Laser Interferometer Space Antenna (LISA) by ESA$\slash$NASA is expected to be launched in 2034~\cite{LISA}. The GW detection via resonant-mass detectors is based on the idea that a GW traveling perpendicular to the mass's axis will produce tidal forces that stretch and contract the mass. If the GW's frequency is close to the resonant frequency of the mass, the deformation of the mass will be detectable. However, such detectors have a very narrow bandwidth because they can only detect frequencies around the resonant frequency, and no GW event has been reported. More GW detection projects based on the four methods above are summarized in ref~\cite{Armstrong2006, Romano2017, Aguiar2010}. \\

The pulsar timing array (PTA) is a program of high-precision timing observations of a widely distributed array of pulsars. A GW induces a Doppler shift, namely a fluctuation in the frequency of a pulse from pulsars to a detector~\cite{Kaufmann1970,Estabrook1975}, and such a fluctuation depends on the position of the GW source, the Earth and the pulsar. In principle, the shift on the frequency can be detected if the corresponding clock (i.e. PTA) has a better stability than the strain of the GWs~\cite{Estabrook1975}. Actually, the shift of the pulse frequency is not measured straightforwardly, but determined from pulse times-of-arrival (ToAs) instead. The ToAs are then compared with predictions of a pulsar timing model. Usually pulsars are observed every few weeks and the longest data sets of observations are a few decades. This implies that PTA data sets are sensitive to GWs with wavelengths from weeks to years, corresponding to ultra-low-frequencies ($10^{-9}$ Hz $\sim10^{-8}$ Hz) GWs.~\cite{Hobbs2017}\\

In analogy to the PTA method, the low-frequency GWs that we are interested in can be detected as well with highly stable clocks and a suitable distance between detectors. Based on this idea, we propose a novel way of GW detection in space, which can be considered complementary to LISA, and which consists of Double Optical Clocks in Space (DOCS) in order to realize Doppler tracking measurement. The two optical clocks in each individual spacecraft can overcome the disadvantages that the Cassini system suffered from.\\

In the traditional Doppler tracking system (e.g. GW detection in the Cassini project), the Earth and a monitored spacecraft act together as freely moving particles, and the distance between them is several astronomical units (AU), which is comparable or larger than the wavelengths of the aimed GWs. From the Earth a radio signal (nominal frequency $\nu_0$) is firstly transmitted to the spacecraft, and then transponded coherently back to the Earth for comparison with a highly stable and precise clock, which is the so-called two way measurement~\cite{Tinto1997}. In principle, a GW propagating through the Earth-spacecraft system causes jitters in relative frequency changes $\delta\nu \slash \nu_0$, and such jittering signal shows up for 3 times in the Doppler response data as functions of time~\cite{Estabrook1975, Armstrong1989}. However, in the two-way method described above the measurement suffers inevitably from two main technical limits. Firstly, in frequency changes of the Doppler response, the noise such as the Earth troposphere, ionosphere and the mechanical vibrations of the ground station comprises a major part in the Doppler signal~\cite{Peng1988, Armstrong1989}. With this, a two one-way measurement method was put forward to cancel the ground-based noise, via mathematically combining the individual Doppler response measured on board the spacecraft and on ground~\cite{Vessot1979, Tinto1996, Tinto2009}. The data process of the two one-way method can improve the sensitivity of GW detection, but its lowest sensitivity ($\sim 10^{-17}$) has no obvious enhancement compared to the two-way measurement~\cite{Tinto2009}. The second limit for Cassini GW detection was the low stability of the H maser on ground, even if it was at the cutting-edge level (Allan deviation $\sigma_y \sim 1\times 10^{-15}$ @ 1000 s) in 1990s.\\

As a GW detection project the proposed DOCS can overcome the main technical difficulties in the traditional earth-space radio link. Due to the two spacecrafts in space, DOCS is expected to avoid the noise from the Earth completely, in other words the frequency fluctuations induced by troposphere, ionosphere and ground-based antenna and transponder are technically removed. Unlike the H maser on ground, two highly stable optical clocks are proposed to be carried in two spacecrafts separately. Moreover, a longer signal integration time $T$ of 2 years (40 days in the case of the Cassini project), which determines the frequency resolution of the signal in frequency domain, is considered in this proposal. All these improvements lead to an optimal sensitivity, which could attain a level of $10^{-19}$.\\

\begin{figure}[htbp]
 \centering
 \includegraphics[width=6cm,height=8cm]{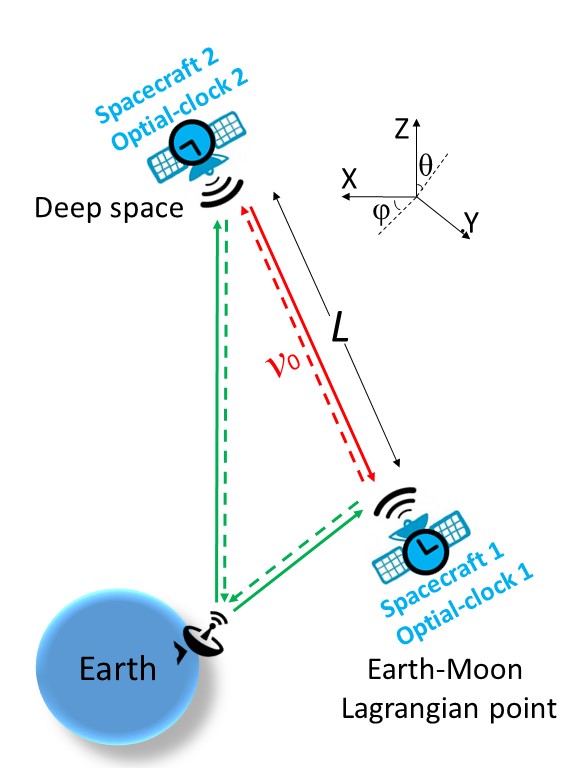}
 \caption{\label{fig:SpaceGWD} Schematic diagram for  DOCS project. Spacecraft 1 and 2 are set to one of the Earth-Moon Lagrangian points and to deep space ($L$ = 1.5 AU), respectively. Each of the spacecrafts  has an optical clock on board. A radio signal with a nominal frequency $\nu_0$ is transmitted from spacecraft 1 (or 2) to spacecraft 2 (or 1), namely using two one-way link. The angles $\theta$ and $\phi$ with respect to the Z-axis and X-axis indicate the direction of the radio link. Meanwhile, the two one-way method between each spacecraft and the Earth act as an auxiliary measure for the signal process.}
 \end{figure}

Figure~\ref{fig:SpaceGWD} shows the scheme of DOCS Doppler tracking system. Two spacecrafts are launched to space, and here we assume that Spacecraft1 (SC1) with optical clock1 and Spacecraf2 (SC2) with optical clock2 are set to one of the Earth-Moon Lagrangian points and deep-space, respectively. The distance $L$ between the two spacecrafts is set to 1.5 AU, which is compatible with the China's Mars Exploration Programme in few years~\cite{Jiang2017}. We assume that both clocks can reach a high stability of $\sigma_y = 4.1\times 10^{-17}$ @ 1000 s, and this stability has recently been reached for a transportable optical clock ~\cite{Koller2017}, which can very likely be available for space application in the near future. The averaging time of 1000 s is comparable to the GW travel time between the two spacecrafts (i.e. 750 s for single trip and 1500 s for return trip). The noise analysis of the two optical clocks is explained in Section IV. Via the two synchronized clocks and radio instruments (e.g. transmitters and receivers) on board, a radio signal (solid and dashed lines in red in Fig~\ref{fig:SpaceGWD}) can be transmitted from SC1 (or 2) to SC2 (or 1), and the Doppler signals as functions of time can be measured simultaneously on the two spacecrafts, namely the two one-way measurement. Meanwhile, two radio links are established between the Earth and the two spacecrafts (solid and dashed lines in green in Fig~\ref{fig:SpaceGWD}). These two channels of Doppler signal are used as an auxiliary measurement to verify the fidelity of the DOCS signal. The signal processing for this Earth-Spacecraft link has already been discussed in review~\cite{Armstrong2006}. Ideally, the three sets of Doppler signals for a GW event can quantitatively determine not only the frequency and the amplitude, but also the propagation direction of the GW.\\

A similar space-based GW detection system was put forward by Kolkowitz et al.~\cite{Kolkowitz2016}, in which two spatially separated, drag-free satellites are set on heliocentric orbit, and each satellite contains an optical lattice atomic clock in order to observe the Doppler shift. Instead of the radio link between the two spacecrafts as in the case of DOCS, they considered an optical laser light sent by conventional optical telescopes compatible with LISA technology. At variance of LISA, the frequency range in their proposed mission is up to 10 Hz, which would bridge the detection gap between space-based and terrestrial optical interferometric GW detectors.\\

Instead of an optical link as used in LISA and in Kolkowitz et al.'s proposal in order to obtain a very high sensitivity, DOCS adopts radio link technique. This choice is a tradeoff between a decent sensitivity and expected technological development in the near future. Moreover, DOCS aims at GWs from the well-known white dwarf binary systems (see more details in Section II), which compensates its relative low sensitivity, equivalently reducing the overall difficulty during observation. A preferred road for DOCS would be to join China's Deep Space Exploration project (e.g. Mars Exploration)~\cite{ChinaDeepSpace}. The ongoing Chinese projects, which already have kept the well-developed radio link technique, can guarantee an expected launch time before 2030 at latest.\\

The goal of this paper is to discuss the feasibility of the DOCS proposal by estimating the sensitivity to the GW signal in the overall Doppler response, and being a preliminary study we will ignore the details of the signal processing instruments (e.g. receiver, transmitter etc.). In Section II we model the Doppler signal on board the two spacecrafts with the two one-way method. In Section III, we calculate the sensitivity of GW detection via DOCS in the interesting frequency range from $10^{-4}$ Hz to $10^{-2}$ Hz, then in Section IV we discuss the noise sources and other candidate settings in DOCS. Finally, we present our conclusions in Section V.\\

\section{Doppler response model in DOCS}
The two one-way Doppler response (or the frequency changes) is induced by both the GW signal and the various noises. We first simulate the GW with a simplified formula~\cite{Estabrook1975}:

\begin{equation}
\label{eq:GW}
h(t) = h_+(t)\rm{cos}(2\phi)+\it h_\times(t) \rm{sin}(2\phi),
\end{equation}
where the GW travels along the Z-axis, and $h_+(t), h_\times(t)$ represent two amplitudes along the two orthogonal axes, X and Y. The Cartesian coordinates (X,Y,Z) with two azimuth angles ($\theta$, $\phi$) are defined in Fig.~\ref{fig:SpaceGWD}.\\

Then the Doppler response including the GW signal recorded on each spacecraft can be written in the similar way of ref~\cite{Piran1986, Tinto1996} as follows
\begin{eqnarray}
\label{eq:S1}
(\frac{\Delta\nu(t)}{\nu_0})_1 && \equiv S_1(t) \nonumber \\
&&= \frac {1}{2}(1-\mu)[h(t-(1+\mu)L) - h(t)] + C_2(t-L) - C_1(t) + B_1(t) \nonumber \\
&&\quad + B_2(t-L) + A_2(t-L) + EL_1(t),
\end{eqnarray}

\begin{eqnarray}
\label{eq:S2}
(\frac{\Delta\nu(t)}{\nu_0})_2 && \equiv S_2(t) \nonumber \\
&&= \frac {1}{2}(1+\mu)[h(t-L) - h(t-\mu L)] + C_1(t-L) - C_2(t) + B_2(t) \nonumber \\
&&\quad  + B_1(t-L) + A_1(t-L) + EL_2(t).
\end{eqnarray}

Eqs.~\ref{eq:S1} and \ref{eq:S2} illustrate the relative changes of frequency with respect to the nominal frequency $\nu_0$ as functions of time measured on SC1 and SC2. Besides the contributions of the GW signals (i.e. the terms depending on $\mu$, $\mu = \rm cos\theta$) to frequency changes, the various noises arising during the radio communication between the two spacecrafts are also taken into account. $C_i$ ($i$ = 1, 2) is associated with the random frequency fluctuations of the optical clocks on the two spacecrafts; here the two clocks are assumed to be synchronized~\cite{Tinto2002, Tinto2009}. $B_i$ ($i$ = 1, 2) represents the noise of the probe `buffeting' caused by forces other than gravity on board the $\rm{SC}_{\it i}$. $A_i$ and $EL_i$ ($i$ = 1, 2) stand for the noise of the amplifiers$\slash$transmitter, and other electronics on board the $\rm{SC}_{\it i}$, respectively. We do not consider the frequency fluctuations caused by the interplanetary plasma, because this noise can be eliminated via high frequency radio link ($K\alpha$ band, 32GHz) and multilink as used in the Cassini mission~\cite{Tinto2009}.\\

As an example, the noise term in Eqs.~\ref{eq:S1} and \ref{eq:S2} `$C_2(t-L) - C_1(t)$' means the sum of the frequency fluctuations of the clock1 on SC1 at moment $t$ and of the clock2 on SC2 $L$ time ago (the speed of light $c$ is set to unity). The minus sign in the example is due to the heterodyne nature of the Doppler measurement~\cite{Tinto2009}. The other terms in Eqs.~\ref{eq:S1} and \ref{eq:S2} can be understood in a similar way. More details for each noise source are discussed in Section IV.\\

In the two one-way measurement method, we combine the two signals $S_1(t)$ and $S_2(t)$ as follows:

\begin{eqnarray}
s(t)&& = S_2(t) - S_1(t-L) \nonumber \\
&& = h(t-L) - \frac {1}{2}(1+\mu)h(t-\mu L) - \frac {1}{2}(1-\mu)h(t-\mu L-2L)  \nonumber \\
&&\quad + 2C_1(t-L) - [C_2(t) + C_2(t-2L)] + [B_2(t) - B_2(t-2L)] \nonumber \\
&&\quad + [A_1(t-L) -A_2(t-2L)] + [EL_2(t) - EL_1(t-L)].
\label{eq:s}
\end{eqnarray}

\section{Results}
In order to calculate the sensitivity of the two one-way signal $s(t)$, we need to know the transfer function of the noise part in $s(t)$, i.e. the fractional frequency one-sided power spectral density (PSD) $n(f)$ of the `non-$h$' terms. This can be obtained by calculating the modulus square of the Fourier transform of each noise term as below:

\begin{equation}
\label{eq:nf}
n(f) = 4S_{C2}\rm{cos}^2(2\it\pi fL) + \rm4\it S_{C\rm1} + \rm4\it S_{B\rm2}\rm{sin}^2(2\it\pi fL) + S_{A\rm1} + \it S_{A\rm2}+ \it S_{EL\rm1} + \it S_{EL\rm2},
\end{equation}

where $S_{Ci}$, $S_{Bi}$, $S_{Ai}$ and $S_{ELi}$ stand for the PSD of noise from optical clock, spacecraft buffeting, amplifiers$\slash$transmitter, and other electronics on SC$i$. Particularly, the noise of the clock2 and the buffeting are treated coherently, while the other noise sources are assumed to be uncorrelated. The PSD for each noise source is listed and discussed in Section IV.\\

The sensitivity of GW detection in DOCS system, which equivalently equals the amplitude ratio of noise and GW with a signal-to-noise ratio of 1, is defined as~\cite{Armstrong2006}

\begin{equation}
\label{eq:SIGMA}
\Sigma(f) = \sqrt{\frac{n(f)B}{q(f)}}.
\end{equation}

In Eq. \ref{eq:SIGMA} $B = 1\slash T = 1.6\times10^{-8}$ Hz is the resolution in the frequency domain, and `$n(f)B$' is the noise power at frequency $f$; $q(f)$ is the transfer function of the averaged power for GW signal in $s(t)$, and we straightforwardly use the expression given in ref~\cite{Tinto2009}.

\begin{figure}[htbp]
 \centering
 \includegraphics[width=13cm,height=10cm]{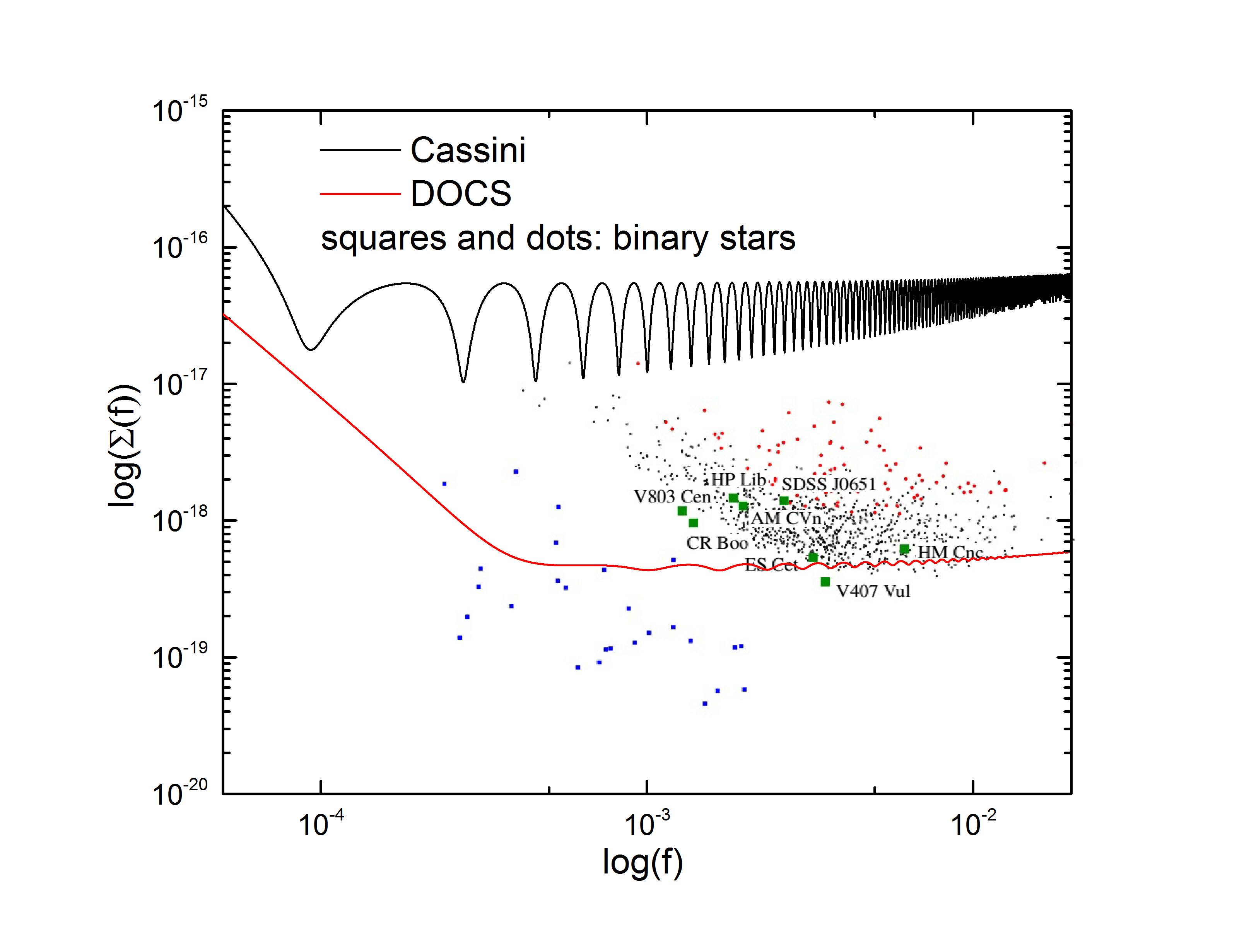}
 \caption{\label{fig:Sigma}Two one-way Doppler response sensitivities and potential GW sources: (1) black curve is the Cassini sensitivity from~\cite{Tinto2009}; (2) red curve is the DOCS sensitivity with estimated noise sources. In the Cassini curve we assume a frequency bandwidth of 2.9$\times$10$^{-7}$ Hz (i.e. a 40-day observation time) and a distance of 5.5 AU~\cite{Tinto1996}. For the DOCS curve we consider as compared to Cassini: (1) two more stable optical clocks ($\sigma_y = 4.1\times10^{-17}@1000s$); (2) a frequency bandwidth of 1.6$\times$10$^{-8}$ Hz due to a longer integration time of 2 years; (3) a distance of $L=$ 1.5 AU. The sensitivity curves of Cassini and DOCS represent the equivalent strain to produce a signal-to-noise ratio of 1. Verification white dwarf binaries are indicated as squares: green ones are labeled with their catalogue names and blue ones are other known binaries~\cite{Nelemans2013,Brown2011,Roelofs2006,Roelofs2010}. Red and black dots are simulated binaries: the strongest 100 are in red and the other 1000 are in black~\cite{Nelemans2004}. The integration time for the binaries is two years~\cite{Seoane2012}.}
 \end{figure}

Figure~\ref{fig:Sigma} shows the calculated GW sensitivities in the frequency range between $5\times 10^{-5}$ Hz and $5\times 10^{-2}$ Hz for DOCS and Cassini project, and some so-called `verification white dwarf binaries' and simulated ones which act as GW sources assuming an integration time of 2 years are plotted as well~\cite{Nelemans2013,Brown2011,Roelofs2006,Roelofs2010,Nelemans2004}. Particularly, the simulated thousand brightest binary systems (dots) are from a population synthesis model for the Galactic population of white dwarf binaries~\cite{Seoane2012}. The black curve corresponds to the sensitivity of the Cassini GW detection system, and the resulting data are reproduced by using the well-measured PSD of the noise sources in the Cassini project as shown in ref~\cite{Tinto2009} (see Table I). The red curve shows the DOCS sensitivity based on the estimated noise PSD at present, which are listed in Table I as well. \\

It is obvious that the sensitivity of DOCS (red curve) is approximately two orders of magnitudes better than Cassini (black curve). With the two years integration, most of the verification binaries (squares) and the simulated double white dwarfs systems (dots) show amplitudes above the sensitivity of DOCS, indicating a high probability of GW detection of these sources, which have periodic wave forms.\\

\section{Discussions}
\subsection{Noise sources}
The sensitivity improvement of DOCS with respect to the conventional radio link system of Cassini mainly originates from three factors. The dominating one is the use of two highly stable optical clocks, which are supposed to provide a much better stability than the H maser used on ground in the Cassini mission. Secondly, the noise is completely removed of troposphere, ionosphere, ground-based antenna and transponder sources, thus all the ground-based noise sources. Finally, DOCS has a longer integration time (2 years), which contributes a factor of $\sqrt{730\slash40} = 4.27$ to the improved sensitivity compared to the Cassini one.\\

\begin{sidewaystable}[h]
\caption{\label{table1} Noise sources for Cassini~\cite{Tinto2009} and DOCS. The details of fractional
frequency one-side power spectral density for the two on board clocks are discussed in Section IV.}
\begin{tabular}{| c | c | c | c | }
    \hline
    \multicolumn{2}{|c|}{GW detection in Cassini} &  \multicolumn{2}{c|}{DOCS} \\
    \hline

    \multirow{2}{2in}{Noise source}       & Fractional frequency one-sided              &  \multirow{2}{2in}{Noise source}       &Fractional frequency one-sided   \\
                                          &power spectral density   &                                        &power spectral density \\
    \hline

    \multirow{3}{2in}{Ground H maser $S_{CE}$\\$\sigma_y = 1.0\times10^{-16}$ @1000s}       & $6.2\times[10^{-28}f + 10^{-33}f^{-1} +10^{-30}]$
    & $^{87}$Sr Optical clock       & $3.38 \times 10^{-30}$   \\
                                                                                           & $ + 1.3\times 10^{-28}f^{2}$
    &on Spacecraft1 $S_{C1}$       & \\

    &                               &$\sigma_y = 4.1\times 10^{-17}$ @1000s                  &\\

    \hline

    \multirow{3}{2in}{Spacecraft clock $S_{CS}$\\}                                        & $5.0\times10^{-27}$ ($10^{-5}\leq f\leq2.0\times10^{-2}$)
    & $^{87}$Sr Optical clock       & $3.38 \times 10^{-30}$   \\
     (a local quartz oscillator and                                                                                    & $2.5\times10^{-25}f$ ($2.0\times10^{-2}\leq f\leq2.0\times10^{-1}$)
    &on Spacecraft2 $S_{C2}$       & \\
     a trapped Hg ions clock)                                                                                        & $10^{-26}f^{-1}$ ($2.0\times10^{-1}\leq f\leq1$)
    &$\sigma_y = 4.1\times 10^{-17}$ @1000s &\\

    \hline

    Spacecraft electronics $S_{ELS}$   &  $7.2\times 10^{-28}f^{2}$  &  Spacecraft electronics $S_{ELS}$  &  $7.2\times 10^{-30}f^{2}$ \\

    \hline

    \multirow{2}{2in}{Ground and onboard \\ Amplifiers $S_{AE} + S_{AS}$}     &  $2.3\times10^{-28} + 4.0\times10^{-25}f$
    & Onboard Amplifiers $S_{AS}$                                           &  $2.3\times10^{-31} + 4.0\times10^{-28}f$ \\
                                 &                                            &                                          &\\

    \hline

    Spacecraft buffeting $S_B$    &  $5.0\times10^{-42}f^{-3}+1.0\times10^{-31}$  &  Spacecraft buffeting $S_B$  &  $5.0\times10^{-44}f^{-3}+1.0\times10^{-33}$\\

    \hline

    Atmosphere $S_T$              & $2.8\times10^{-28}f^{-2/5}$             &  Atmosphere $S_T$             &                  0                    \\

    \hline

    Transponder $S_{TR}$          & $1.6\times10^{-26}f$                    &  Transponder $S_{TR}$         &                  0                     \\

    \hline

    Ground electronics $S_{ELE}$  & $6.3\times10^{-27}f^{2}$                & Ground electronics $S_{ELE}$  &                  0                     \\

    \hline

\end{tabular}
\end{sidewaystable}

Table I lists the fractional frequency one-sided PSD of all various noise sources discussed above, where the column `Cassini' refers to ref~\cite{Tinto2009}. For `Cassini', the H maser clock has an Allan deviation $\sigma_y = 1.0\times10^{-16}$ @1000s~\cite{Tinto1996,Tinto2009} and its space clock is  proposed with a combination of a local quartz oscillator and a trapped Hg ions clock~\cite{Tinto2009}; while for DOCS, considering the tradeoff between the recent advances of optical clocks and the technology readiness level for space applications, we assume to use the parameters of a transportable optical clock in space as follows $\sigma_y = 4.1 \times 10^{-17}@1000s$ ($\sigma_y(\tau) = 1.3 \times 10^{-15}\slash\sqrt{\tau}$, $\tau$ is the average time) for the two clocks~\cite{Koller2017}. In order to calculate the fractional frequency one-sided PSD $S_{Ci}$ ($i$ = 1, 2) for the transportable optical clock, we use the following relation~(see \cite{Barnes1971}):

\begin{equation}
\label{eq:Allan}
\sigma^2_{y}(\tau) = 2\int_0^\infty S_{y}(f)\frac{\rm sin^4(\pi\tau \it f)}{(\pi\tau f)^2}df,
\end{equation}

where $\sigma_y(\tau)$ is known and $S_{y}(f) = S_{Ci}(f)$ is determined using Eq. \ref{eq:Allan} (`$S_{y}(f)$' is a conventional term for the fractional frequency one-sided PSD). The power spectrum of most atomic clocks can be described with a simple model for $S_{y}(f) = \sum_{\alpha=-2}^2h_{\alpha}f^{\alpha}$, where `$f^{\alpha}$' ($\alpha$ = -2, -1, 0, 1 and 2) reflects the various contributions of noise in the clock system (i.e. random walk frequency noise, flicker frequency noise, white frequency noise, flicker phase noise, and white phase noise~\cite{Barnes1971}). The behaviour of $1\slash\sqrt{\tau}$ for $\sigma_y(\tau)$ indicates that the white frequency noise dominates~\cite{Santorini2016} and therefore $S_y(f) = h_0$. From Eq. \ref{eq:Allan} we get $h_0 = 2\tau\times \sigma^2_y(\tau)= 3.38 \times 10^{-30}$ $\rm Hz^{-1}$.\\

The second important contribution to the noise is due to the amplifiers and transmitters on the spacecrafts~\cite{Tinto2009}. In order to estimate it we rescaled the noise PSD $S_{Ai}$ ($i$ = 1, 2) to $10^{-3}$ of its original value, because $S_{Ai}$ of Cassini includes both the noise originating on ground and on board the spacecraft, which is highly overestimated for DOCS as Cassini technology is more than twenty years old. Considering the low noise amplifiers available today, our assumption on the rescaling factor for the noise should be realistic~\cite{Bryerton2013}.\\

For buffeting noise PSD $S_{Bi}$ ($i$ = 1, 2), we expect at least 100 times improvement for DOCS. Especially, SC1 is located in one of the Earth-Moon Lagrangian points, which might lead to even better stability than the conservative assumption made here.\\

Since the onboard clock provides the fundamental frequency and timing reference for the main onboard radio instrumentation, and determines the stability of the microwave signal transmitted to the ground~\cite{Tinto2009}, the electronics noise PSD $S_{ELi}$ ($i$ = 1, 2) on board of DOCS plays an important role and we assume that we can gain a further 100 times improvement compared to that of Cassini as indeed much more stable optical clocks will be used.\\

\subsection{Estimated improvements on DOCS in the next decade}
The stability of the optical clocks and the suppression of various noise contributions in the above calculations are conservative and should be feasible at present. Here, we discuss possible further improvements for the main parameters, which could be achieved in space in the next decade. For instance, if we take one of the cutting-edge optical lattice clocks ($\sigma_y(\tau) = 3.1\times10^{-17}\slash\sqrt{\tau}$~\cite{Campbell2017}) as the clocks to be used in the two spacecrafts with a stability of $9.8\times10^{-19}$ at $\tau$ = 1000 s, and we suppress the noise of amplifiers on board the spacecrafts, buffeting and other electronics by at least one order of magnitude, we could further improve the overall sensitivity of DOCS (still assuming two years integration) by at least one order of magnitude with respect to its original sensitivity as discussed above.\\

Based on these further assumed improvements of clocks and electronics, one could also envisage using a much shorter observation time (about one week to one month) without sacrificing the original sensitivity ($10^{-19}$). Thus, GWs from some predicted, non-periodic sources could be detected, including emission from coalescence of supermassive black holes, rotational instabilities in protoneutron stars~\cite{Piro2011,Corsi2009}, and black-hole accretion disk instabilities~\cite{Kiuchi2011}. The detection of GWs from such sources using an improved DOCS is however beyond the scope of this paper, and it needs to be studied in details. We shall come back to this issue in a following paper.\\

\subsection{Other possible configurations for DOCS}
In the above discussion of DOCS we assumed to set one spacecraft (SC1) at one of the Earth-Moon Lagrangian points, and the other (SC2) to the deep space. Of course, other configurations for SC1, e.g. on a geostationary orbit or at one of Sun-Earth Lagrangian points, could be envisaged as above. Of course such choices will need further investigations by considering both technical and budgetary issues.\\

We find that the calculated sensitivity does not depend much on the value of $L$ in the range from 1 to 10 AU. In this work we assumed $L$ to be 1.5 AU, as we might expect a possible cooperation with the coming China's Mars Exploration Programme. Moreover, without decreasing the sensitivity the distance between the two spacecrafts in DOCS could also be increased, e.g. to 5 AU (similar to the case of Cassini), and therefore lower frequency GWs would be detectable.\\

\subsection{Tests of deviation from General Relativity}

Einstein's general relativity has so far passed all experimental and observational tests~\cite{Will2006} including GW detection~\cite{Abbott2016}. However, for explaining dark matter and dark energy some alternative gravity theories have been proposed. To detect such deviations in general relativity is the most important scientific goal besides GW detection. DOCS with its high sensitivity clocks would be well suited to detect small deviation in GR, e.g. using the gravitational redshift phenomenon~\cite{Vessot1980,Delva2015,Danzmann2015}. A more detailed study on such possibilities will be presented in a future paper.\\

\subsection{Technical issues to be solved}

Two key techniques are indispensable in DOCS: highly stable clocks and low noise RF amplifiers. The up-to-date transportable optical clocks considered in this work show Allan deviations of $10^{-17}$ at $\tau$ = 1000 s. Over the last decades optical clocks have exhibited a significant improvement, with a factor of $10^{-3}$ for the Allan deviation per decade~\cite{Gill2011}. Based on these improvements an optical clock with an Allan deviation of $9.8\times10^{-19}$ at $\tau$ = 1000 s~\cite{Campbell2017} should be realistic, although this assumption is a conservative estimate for the sensitivity improvement in the next decade, as discussed in subsection B.\\

Low noise amplifiers for high frequency radio (K$\alpha$-band) link as proposed to be used in DOCS is the other important issue. Taking the Cassini experiment as a reference, in which a K$\alpha$-band Travelling Wave Tube Amplifier (TWTA) on board the spacecraft had an output power of 10 W for a distance of 5 AU~\cite{Tinto2009}, therefore for the distance of 1.5 AU as in DOCS, an amplifier roughly needs an output power of $10\times(1.5\slash5)^2 W \thickapprox 1 W$. This amplifier can be easily satisfied, e.g. today the most advanced amplifier in space for K$\alpha$-band communication can reach an output of 250 W with a bandwidth of 2.9 GHz~\cite{Ehret2014,Ayllon2015}. However, in Cassini project a 4 meter diameter antenna on board the spacecraft, and especially a 34 meter diameter antenna receiver on ground were used to keep a decent low thermal noise level~\cite{Tinto2009}. It is not yet clear for DOCS if we could achieve a similar or even lower noise level by using similar antennas on board of the two spacecrafts (i.e. a 4 meter diameter antenna on each spacecraft). We will thus investigate in more details the noise level by considering both the use of two antennas (e.g. with 4 meter diameter) on board the spacecrafts and the absence of the ground-based noise.\\

Should it turn out to be too difficult to use a radio link technique, an optical link could be envisaged as an alternative for DOCS. Besides Kolkowitz et al.'s proposal, another optical-link-based idea of `a xylophone interferometer detector of gravitational radiation' was put forward as well~\cite{Tinto1998}. This latter proposal is based on two spacecrafts tracking each other via coherent laser light. After combining two one-way data and selecting proper Fourier components, the frequency fluctuations introduced by the lasers can be reduced by several orders of magnitudes, and the remaining GW signal at these low noise frequencies can be detected. Either using radio- or optical-link techniques, the two-spacecraft-based configurations like DOCS, Kolkowitz et al.'s proposal and the xylophone interferometer detector could be complementary to each other.\\

\section{Conclusions}
We have proposed a novel Doppler tracking system with Double Optical Clock in Space for gravitational wave (mainly for periodic waveforms) detection. Compared to the conventional radio link used in the Cassini project, the new proposed system including two space-based spacecrafts avoids the frequency fluctuations from troposphere, ionosphere and ground-based antenna and transponder. Moreover, two much more stable optical clocks on board the spacecrafts are taken into consideration in order to substantially increase the detection sensitivity, and the longer signal observation time of 2 years improves the detection resolution in the considered frequency domain. By taking all the noise sources into account, we finally obtain a signal sensitivity of $5\times10^{-19}$ in the frequency range from $10^{-4}$ Hz to $10^{-2}$ Hz. At such a sensitivity DOCS is expected to detect GWs from the galactic white dwarf binaries. Besides GW detection, general relativity test (e.g. gravitational redshift) can also be performed at the same time. Considering the foreseeable technological improvements including optical clocks, DOCS could also detect other sources of GWs (e.g. GWs from coalescence of supermassive black holes, protoneutron stars and black-hole accretion disk instabilities) making the science case for DOCS even more appealing.\\

\section{acknowledgement}
We thank Dr. Massino Tinto (University of California San Diego) for the important discussions on RF hardware in space. This work was funded by the project `Research on space-time structure of the universe by space-based optical clocks, XDA15007702', which is financially supported by National Space Science Center, The Chinese Academy of Sciences.
\bibliographystyle{unsrt}
\bibliography{References}

\begin{thebibliography}{10}

\bibitem{Einstein1916}
A.~Einstein.
\newblock N$\rm\ddot{a}$herungsweise integration der feldgleichungen der
  gravitation.
\newblock {\em Sitzungsberichte der K$\rm\ddot{o}$niglich Preussischen Akademie
  der Wissenschaften Berlin. part 1: 688-696}, 1916.

\bibitem{Einstein1918}
A.~Einstein.
\newblock $\rm\ddot{U}$ber gravitationswellen.
\newblock {\em Sitzungsberichte der K$\rm\ddot{o}$niglich Preussischen Akademie
  der Wissenschaften Berlin. part 1: 154-167}, 1918.

\bibitem{Einstein1937}
A.~Einstein and N.~Rosen.
\newblock On gravitational waves.
\newblock {\em J Franklin Inst}, 223:43, 1937.

\bibitem{Moore2014}
C.~J. Moore, R.~H. Cole, and C.~P.~L. Berry.
\newblock Gravitational-wave sensitivity curves.
\newblock {\em Classical and Quantum Gravity}, 32:1, 2014.

\bibitem{Tinto2002}
M.~Tinto.
\newblock Spacecraft radio occultations using multiple doppler readouts.
\newblock {\em Radio Sci.}, 37:1045, 2002.

\bibitem{Armstrong2006}
J.W. Armstrong.
\newblock Low-frequency gravitational wave searches using spacecraft doppler
  tracking.
\newblock {\em Living Reviews in Relativity}, 9:1, 2006.

\bibitem{Abbott2016}
B.~P. Abbott and et~al.
\newblock Observation of gravitational waves from a binary black hole merger.
\newblock {\em Phys. Rev. Lett.}, 116:061102, 2016.

\bibitem{Estabrook1975}
F.B. Estabrook and H.D. Wahlquist.
\newblock Response of doppler spacecraft tracking to gravitational radidation.
\newblock {\em Gen. Relativ. Gravit.}, 6:439, 1975.

\bibitem{Seoane2012}
P.~Amaro-Seoane and et~al.
\newblock Low-frequency gravitational-wave science withe lisa-ngo.
\newblock {\em Class Quantum Grav.}, 29:124016, 2012.

\bibitem{Seoane2017}
P.~Amaro-Seoane and et~al.
\newblock Laser interferometer space antenna.
\newblock {\em arXiv:1702.00786}, 2017.

\bibitem{Luo2016}
J.~Luo and et~al.
\newblock A space-borne gravitational wave detector.
\newblock {\em Class Quantum Grav.}, 33:035010, 2016.

\bibitem{Weber1961}
J.~Weber.
\newblock {\em General Relativity and Gravitational Waves}.
\newblock Interscience Publishers Inc., New York, 1961.

\bibitem{Aguiar2010}
O.~D. Aguiar.
\newblock The past, present and future of the resonant-mass gravitational wave
  detectors.
\newblock {\em arXiv:1009.1138}, 2010.

\bibitem{Backer1986}
D.C. Backer and R.~W. Hellings.
\newblock Pulsar timing and general relativity.
\newblock {\em Annu Rev Astro Astrophys}, 24:537, 1986.

\bibitem{Cassini}
\url{https://www.nasa.gov/mission_pages/cassini/main/index.html}.

\bibitem{LISA}
\url{https://lisa.nasa.gov/}.

\bibitem{Romano2017}
J.~D. Romano and N.J. Cornish.
\newblock Detection methods for stochastic gravitational-wave backgrounds: a
  unified treatment.
\newblock {\em Living Reviews in Relativity}, 20:2, 2017.

\bibitem{Kaufmann1970}
W.~J. Kaufmann.
\newblock Redshift fluctuations arising from gravitational waves.
\newblock {\em Nature}, 227:157, 1970.

\bibitem{Hobbs2017}
G.~Hobbs and S.~Dai.
\newblock A review of pulsar timing array gravitational wave research.
\newblock {\em arXiv:1707.01615}, 2017.

\bibitem{Tinto1997}
M.~Tinto and J.~W. Armstrong.
\newblock {\em Narrow-Band Searches for Gravitational Radiation With Spacecraft
  Doppler Tracking}.
\newblock Proceedings of the 29th Annual Precise Time and Time Interval Systems
  and Applications Meeting, Long Beach, California, 1997.

\bibitem{Armstrong1989}
J.W. Armstrong.
\newblock Advanced doppler tracking experiments.
\newblock {\em In NASA, Relativistic Gravitational Experiments in Space},
  page~70, 1989.

\bibitem{Peng1988}
T.K. Peng, J.W. Armstronga, J.C. Breidenthal, F.F. Donivan, and N.C. Ham.
\newblock Deep space network enhancement for the galileo mission to jupiter.
\newblock {\em Acta Astronautica}, 17:321, 1988.

\bibitem{Vessot1979}
R.F.C. Vessot and M.W. Levine.
\newblock A test of the equivalence principle using a space-borne clock.
\newblock {\em Gen. Relativ. Gravit.}, 10:181, 1979.

\bibitem{Tinto1996}
M.~Tinto.
\newblock Spacecraft doppler tracking as a xylophone detector of gravitational
  radiation.
\newblock {\em Phys. Rev. D}, 53:5354, 1996.

\bibitem{Tinto2009}
M.~Tinto, G.J. Dick, J.D. Prestage, and J.W. Armstrong.
\newblock Improved spacecraft radio science using an on-board atomic clock:
  Application to gravitational wave searches.
\newblock {\em Phys. Rev. D}, 79:102003, 2009.

\bibitem{Jiang2017}
X.~Jiang, B.~Yang, and S.~Li.
\newblock Overview of china’s 2020 mars mission design and navigation.
\newblock {\em Astrodynamics}, page~1, 2017.

\bibitem{Koller2017}
S.~B. Koller, J.~Grotti, St. Vogt, A.~Al-Masoudi, S.~Dörscher, S.~Häfner,
  U.~Sterr, and Ch. Lisdat.
\newblock Transportable optical lattice clock with $7 \times 10^{−17}$
  uncertainty.
\newblock {\em Phys. Rev. Lett.}, 118:073601, 2017.

\bibitem{Kolkowitz2016}
S.~Kolkowitz, I.~Pikovski, N.~Langellier, M.~D. Lukin, R.~L. Walsworth, and
  J.~Ye1.
\newblock Gravitational wave detection with optical lattice atomic clocks.
\newblock {\em Phys. Rev. D}, 94:124043, 2016.

\bibitem{ChinaDeepSpace}
\url{http://www.cnsa.gov.cn/n6443408/n6465652/n6465653/c6768527/content.html}.

\bibitem{Piran1986}
T.~Piran, E.~Reiter, W.G. Unruh, and R.F.C. Vessot.
\newblock Filtering of spacecraft doppler tracking data and detection of
  gravitational radiation.
\newblock {\em Phys. Rev. D}, 34:984, 1986.

\bibitem{Nelemans2013}
G.~Nelemans.
\newblock Galactic binaries with elisa.
\newblock {\em 9th LISA Symposium(Paris)ASP Conference Series}, 467:27, 2012.

\bibitem{Brown2011}
W.~R. Brown, M.~Kilic, J.~J. Hermes, C.~A. Prieto, S.~J. Kenyon, and D.~E.
  Winget.
\newblock A 12 minute orbital period detached white dwarf eclipsing binary.
\newblock {\em ApJ}, 737:23, 2011.

\bibitem{Roelofs2006}
G.~H.~A. Roelofs, P.~J. Groot, G.~Nelemans, T.~R. Marsh, and D.~Steeghs.
\newblock Kinematics of the ultracompact helium accretor am canum venaticorum.
\newblock {\em MNRAS}, 371:1231, 2006.

\bibitem{Roelofs2010}
G.~H.~A. Roelofs, A.~Rau, T.~R. Marsh, D.~Steeghs, P.~J. Groot, and
  G.~Nelemans.
\newblock Spectroscopic evidence for a 5.4 minute orbital period in hm cancri.
\newblock {\em ApJ}, 711:138, 2010.

\bibitem{Nelemans2004}
G.~Nelemans, L.~R. Yungelson, and S.~F.~Portegies Zwart.
\newblock Short-period am cvn systems as optical, x-ray and gravitational-wave
  sources.
\newblock {\em MNRAS}, 349:181, 2004.

\bibitem{Barnes1971}
J.A. Barnes, A.R. Chi, L.S. Cutler, D.J. Healey, D.B. Leeson, T.E. McGuniga,
  J.A. Mullen, W.L. Smith, R.L. Sydnor, R.F.C. Vessot, and G.M.R. Winkler.
\newblock Characterization of frequency stability.
\newblock {\em IEEE Trans. Instrum. Meas.}, 20:105, 1971.

\bibitem{Santorini2016}
G.~Santorini.
\newblock Optical fiber links for time \& frequency metrology.
\newblock
  \url{https://physique.cuso.ch/fileadmin/physique/cours_commun/cuso-2016_course7_giorgiosantarelli.pdf}.

\bibitem{Bryerton2013}
E.W. Bryerton, M.~Morgan, and M.W. Pospieszalski.
\newblock Ultra low noise cryogenic amplifiers for radio astronomy.
\newblock {\em Radio and Wireless Symposium (RWS), IEEE}, page 358, 2013.

\bibitem{Campbell2017}
S.~L. Campbell, R.~B. Hutson, G.~E. Marti, A.~Goban, N.~Darkwah Oppong, R.~L.
  McNally, L.~Sonderhouse, J.~M. Robinson, W.~Zhang, B.~J. Bloom, and J.~Ye.
\newblock A fermi-degenerate three-dimensional optical lattice clock.
\newblock {\em Science}, 358:90, 2017.

\bibitem{Piro2011}
A.~L. Piro and C.~D. Ott.
\newblock Supernova fallback onto magnetars and propeller-powered supernovae.
\newblock {\em The Astrophysical Journal}, 736:108, 2011.

\bibitem{Corsi2009}
A.~Corsi and P.~Meszaros.
\newblock Gamma-ray burst afterglow plateaus and gravitational waves:
  Multi-messenger signature of a millisecond magnetar?
\newblock {\em The Astrophysical Journal}, 702:1171, 2009.

\bibitem{Kiuchi2011}
K.~Kiuchi, M.~Shibata, P.~J. Montero, and J.~A. Font.
\newblock Gravitational waves from the papaloizou-pringle instability in
  black-hole-torus systems.
\newblock {\em Phys. Rev. Lett.}, 106:251102, 2011.

\bibitem{Will2006}
C.~M. Will.
\newblock The confrontation between general relativity and experiment.
\newblock {\em Living Rev. Relativity}, 6:3, 2006.

\bibitem{Vessot1980}
R.~F.~C. Vessot, M.~W. Levine, E.~M. Mattison, E.~L. Blomberg, T.~E. Hoffman,
  G.~U. Nystrom, B.~F.~F. Decher, P.~B. Eby, C.~R. Baugher, J.~W. Watts, D.~L.
  Teuber, and F.~D. Wills.
\newblock Test of relativistic gravitation with a space-borne hydrogen maser.
\newblock {\em Phys. Rev. Lett.}, 45:2081, 1980.

\bibitem{Delva2015}
P.~Delva, A.~Hees, S.~Bertone, E.~Richard, and P.~Wolf.
\newblock Test of the gravitational redshift with stable clocks in eccentric
  orbits: application to galileo satellites 5 and 6.
\newblock {\em Class. Quantum Grav.}, 32:232003, 2015.

\bibitem{Danzmann2015}
K.~Danzmann for~the LISA Pathfinder~Team and the~eLISA Consortium.
\newblock Lisa and its pathfinder.
\newblock {\em Nature Physics}, 11:613, 2015.

\bibitem{Gill2011}
P.~Gill.
\newblock When should we change the definition of the second?
\newblock {\em Phil. Trans. R. Soc. A}, 369:4109, 2011.

\bibitem{Ehret2014}
P.~Ehret, A.~Laurent, and E.~Bosch.
\newblock Broadband traveling wave tubes in k$\alpha$- and ku-band.
\newblock {\em IEEE International Vacuum Electronics Conference 2014}, 2014.

\bibitem{Ayllon2015}
N.~Ayllon.
\newblock Microwave high power amplifier technologies for space-borne
  applications.
\newblock {\em Wireless and Microwave Technology Conference (WAMICON), 2015
  IEEE 16th Annual}, 2015.

\bibitem{Tinto1998}
M.~Tinto.
\newblock Spacecraft to spacecraft coherent laser tracking as a xylophone
  interferometer detector of gravitational radiation.
\newblock {\em Phys. Rev. D}, 58:102001, 1998.

\end{thebibliography}

\end{document}